# A City Is a Complex Network

Bin Jiang

Faculty of Engineering and Sustainable Development, Division of GIScience
University of Gävle, SE-801 76 Gävle, Sweden
Email: bin.jiang@hig.se

*(Draft: August 2015, Revision: December 2015, January, February 2016)*

**Abstract**
A city is not a tree but a semi-lattice. To use a perhaps more familiar term, a city is a complex network. The complex network constitutes a unique topological perspective on cities and enables us to better understand the kind of problem a city is. The topological perspective differentiates it from the perspectives of Euclidean geometry and Gaussian statistics that deal with essentially regular shapes and more or less similar things. Many urban theories, such as the Central Place Theory, Zipf's Law, the Image of the City, and the Theory of Centers can be interpreted from the point of view of complex networks. A livable city consists of far more small things than large ones, and their shapes tend to be irregular and rough. This chapter illustrates the complex network view and argues that we must abandon the kind of thinking (mis-)guided by Euclidean geometry and Gaussian statistics, and instead adopt fractal geometry, power-law statistics, and Alexander's living geometry to develop sustainable cities.

## 1. Introduction

A city is not a tree but a complex network. Implicit in Alexander's earlier works (e.g., Alexander 1965), this insight on city is a foundation for the Theory of Centers (Alexander 2002–2005). According to the theory, a whole consist of numerous, recursively defined centers (or *sub-wholes*) that support each other. A city is a whole, as is a building, or a building complex. The centers and their nested, intricate relationship constitute a complex network (see below for further discussion). The complex network offers a unique perspective for better understanding the kind of problem a city is (Jacobs 1961). Based on the premise that a whole is greater than the sum of its parts, complexity science has developed a range of tools, such as complex networks (Newman et al. 2006) and fractal geometry (Mandelbrot 1982), for enhancing our understanding of complex phenomena. Unlike many other pioneers in the field, Alexander's contribution to complexity science began with creation or design of beautiful buildings. The Theory of Centers, or living geometry, is much more broad and profound than fractal geometry. Living geometry aims for creation (Mehaffy and Salingaros 2015), while fractal geometry is mainly for understanding. Creation or design is the highest status of science. This chapter will elaborate on the network city view and how its advance significantly contributes to a better understanding of fractal structure and nonlinear dynamics of cities. I will begin with hierarchy within, and among, a set of cities, then illustrate beauty and images emerging from a complex network of centers, and end up with further discussions on fractal geometry and living structure for sustainable urban design.

## 2. Hierarchy within, and among, cities

A city is not a complex network seen from individual street segments or junctions. This is because both street segments and junctions have more or less similar degrees of connectivity (approximately four), very much like a regular or random network. However, a city is a complex network seen from individual streets. The streets are created from individual street segments with the same names or good continuity; so-called *named* and *natural streets* (Jiang and Claramunt 2004, Jiang et al. 2008).

Unlike street segments that are more or less similar, there are all kinds of streets in terms of lengths or degrees of connectivity. The topological view helps develop new insights into cities. To illustrate, let us look at the street network of the historic part of the city Avignon in France. The network comprises 341 streets, which are put into six hierarchical levels based on the head/tail breaks, a classification scheme, as well as a visualization tool, for data with a heavy-tailed distribution (Jiang 2013a, Jiang 2015a). Given the set of streets as a whole, we break it into the head for those above the mean and the tail for those below the mean, and recursively continue the breaking process of the head until the notion of far more less-connected streets than well-connected ones is violated; the head/tail breaks process can be stated as a recursive function as follows.

```
Recursive function Head/tail Breaks:
    Break a whole into the head and the tail;
    // the head for those above the mean
    // the tail for those below the mean
    While (head <= 40%):
        Head/tail Breaks (head);
End Function
```

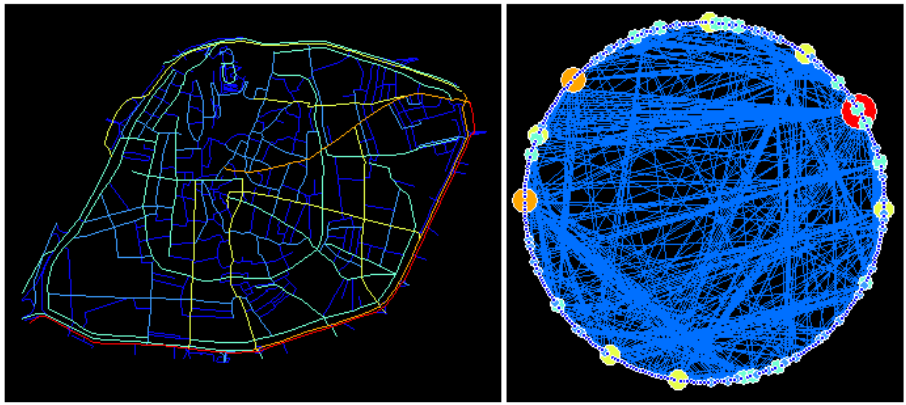

Figure 1: (Color online) Hierarchy of the street network of Avignon, and its connectivity graph both showing far more less-connected streets than well-connected ones
(Note: The hierarchy is visualized by the spectral color with blue for the least-connected streets and red for the most-connected ones. The 341 streets and their 701 relationships become the 341 nodes and 701 links of the connectivity graph.)

The head/tail breaks enables us to see the parts and the inherent hierarchy. The resulting hierarchy is visualized using the spectral color, with blue for the least-connected streets and red for the most-connected ones (Figure 1a). The 341 streets and their 701 relationships (intersections) are converted respectively into the nodes and links of a connectivity graph (Figure 1b). The connected graph is neither regular nor random, but a small-world network – a middle status between the regular and random counterparts (Jiang and Claramunt 2004, Watts an Strogatz 1998). The ring-like visualization shows the connectivity graph with a striking hierarchy of far more small nodes than large ones, with node sizes indicating the degrees of connectivity. Networks with this scaling hierarchy have an efficient structure, commonly known as *scale-free networks* (Barabási and Albert 1999). Both small world and scale free are two distinguished properties of complex networks. A complex network is highly efficient, both locally and globally, inherited respectively from the regular and random counterparts. How is a complex network developed? What are the underlying mechanisms of complex networks? How do we design a complex network of high efficiency? These questions are design oriented, with far-reaching implications for architectural design and city planning. Inspired by Alexander's works (Alexander 2002–2005), a theory of network city (Salingaros 2005) has already been developed for dealing with various urban-design issues.

Not only a city but also a set of cities (or human settlements, to be more precise) is a complex network.

All cities in a large country tend to constitute a whole, as formulated by Zipf's Law (Zipf 1949) and in the Central Place Theory (Christaller 1933, 1966). According to Zipf's Law, city sizes are inversely proportional to their rank. Statistically, the first largest city is twice as big as the second largest, three times as big as the third largest, and so on. Zipf's Law is a statistical law on city-size distribution, and it does not say anything about how the cities are geographically distributed. The geographical distribution of cities is captured by the Central Place Theory. Cities in a country or region tend to be distributed in a nested manner, i.e., each city acts as a central place, providing services to the surrounding areas. Conversely, small cities tend to support large ones, which further support even larger ones in a nested manner. The Central Place Theory is about a network of cities or human settlements that constitute a scaling hierarchy. The underlying network structure formulated by the Central Place Theory resembles the structure of a whole, in which recursively defined centers tend to support each other (Alexander 2002–2005, Jiang 2015b). In this regard, cities in a country or region can be considered to be a living structure.

### 3. Beauty and image out of complex networks

Alexandrian living structure is a *de facto* complex network of numerous centers. The centers are recursively defined, which means that a center contains smaller centers and is contained within larger centers. Besides the nested, intricate relationships among the numerous centers, they tend to support each other to constitute a whole. In this context, wholeness, as defined by Alexander (2002–2005), is considered to be a global structure or life-giving order emerging from the whole as a complex network of the centers. This complex-network view of whole captures the mathematical model of wholeness as part of the Theory of Centers, and enables us to compute the degrees of wholeness or beauty (Jiang 2015b). Using Google's PageRank algorithm, beautiful centers are defined as those to which many beautiful centers point. This definition of beautiful centers is recursive, and computation of the degree of beauty is achieved through an iterative process until a convergence is reached. Eventually, each center is assgined to a degeee of beauty between 0 and 1. The degree of beauty of the whole can be measured by the ht-index, a head/tail breaks-induced index; the higher the ht-index, the more beautiful the whole. Let us illustrate the computation using the Alhambra plan as a working example at a building complex scale.

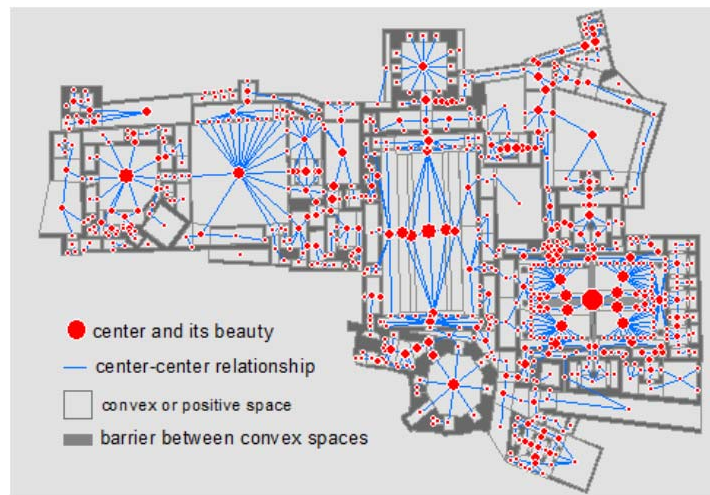

Figure 2: (Color online) The complex network of the centers with the plan of Alhambra
(Note: The degrees of beauty are calculated using Google's PageRank algorithm; the bigger the dots, the more beautiful the centers.)

The Alhambra is probably the most beautiful building complex in the world. It possesses many of the 15 geometric properies such as levels of scale, strong centers, thick boundaries, and local symmetries. Seen from its plan, the most distinguished property is local symmetries. The plan does not look globally symmetric, but the numerous local symmetries make it unique and beautiful. Let us focus on the Alhambra plan that is partitioned into 725 convex spaces, each of which acts as a center. Most of the centers are related to surrounding centers, as long as there is no barrier between them. This makes 880 relationships in total. There are a few isolated centers that do not contribute to the whole. The 880 relationships are directed from the peripheral small spaces to the central large spaces. Figure 4 shows the result, in which the dots indicate the degrees of beauty; the bigger the dots, the more beautiful the centers. It should be noted that there are 13 centers hidden or embedded in the network: One as the whole, three subwholes, and nine subwholes of the three subwholes.

The living structure has deep implications for understanding the city structure from a cogntive perspective. In this connection, the image of the city (Lynch 1960) is another classic in the field of urban design. A large body of literature has been produced over the past 50 years. Much of of the literature focuses on human internal representation, or how do mental images of a city vary from person to person? In fact, it is the city's external representation, or the city itself, or the living structure, that makes a city imageable or legible (Jiang 2013b). To be more precise, the largest, the most-connected, or the most meaningful constitute part of a mental image of the city. Among the five city elements (paths, edges, districts, nodes, and landmarks), only landmarks capture the true sense of scaling or living structure. The image of the Alhambra plan consists of three subwholes: The left, middle, and right. Each of these comprises three further subwholes. Among the many other centers, the most beautiful one, or the one with the most dense local symmetries, tends to shape our image of the building complex.

## 4. Fractal and living structures

The topological perspective differentiates it fundamentally from the perspectives offered by Euclidean geometry and Gaussian statistics. Euclidean geometry aims for measuring regular shapes, and Gaussian statistics aims for analyzing more or less similar things. These two mathematical tools show some constraints while dealing with complexity of the world. Instead of more or less similar things and regular shapes, there are far more small things than large ones, and irregular shapes. To put them in perspective, Euclidean geometry aims for measurement or scale, while fractal geometry aims for scaling or the scaling pattern of far more small things than large ones. Gaussian statistics aims for average things, while power-law statistics aims for outliers. Events of a small probability in Gaussian statistics are impossible, whereas events of a small probability in power-law statistics are highly improbable or vital. To a great extent, Euclidean and fractal geometries complement each other, and one cannot stand without another. This is because one must measure all things under the framework of Euclidean geometry to recognize scaling. However, our thinking in architecture and urban design is very much dominated by Euclidean and Gaussian thinking. For example, to characterize a tree, we tend to only measure its height, rather than all its branches. To illustrate, let us examine two patterns shown in Figure 3.

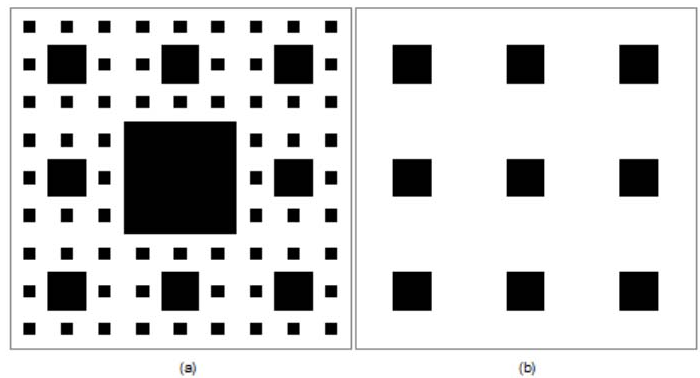

Figure 3: Fractal versus Euclidean patterns
(Note: The left pattern appears appears metaphorically in traditional buildings, in the sense of all scales involved rather than of precisely the same pattern, whereas the right pattern is pervasively seen in modern buildings metaphically and in terms of precisely the same pattern.)

The square of one unit is cut into nine congruent squares, and the middle one is taken away. The same procedure is recursively applied to the remaining eight squares again and again, until we end up with the pattern commonly known as Sierpinski carpet (Figure 3a). This particular carpet of three iterations comprises one square of scale 1/3, eight squares of scale 1/9, and 64 squares of scale 1/27. A Sierpinski carpet is hardly seen in reality, but it helps illustrate some unique properties shared by the real-world patterns, referring to not only those in nature but also those emerging in cities and buildings. First, a pattern recurs again and again at different scales, known as self-similarity. Second, there are multiple scales, rather than just one. It is essentially these two properties that differentiate the left pattern from the right one in Figure 3. It is important to note that the right pattern is with nine squares, which are disconnected each other. However, all the squares of the left pattern are connected each other, according to Gestalt psychology (Köhler 1947). The largest square is supported by the eight middle-sized squares, each of which is further supported by the eight smallest squares. This support relationship is very much similar to the framework of the Central Place Theory.

Unfortunately, modern architecture has been deadly misguided by Euclidean geometry and Gaussian thinking towards so-called geometric fundamentalism (Mehaffy and Salingaros 2006). Geometric fundamentalism worships simple and large-scale Euclidean shapes, such as cylinders and cubes, so removes small scales and ornament. However, scaling laws tell us that all scales ranging from the smallest to the largest (to be more precise, many smallest, a very few largest, and some in between) are essential for scaling hierarchy and for human beings. This scaling hierarchy appears pervasively in traditional buildings such as temples, mosques, and churches, yet has been removed from contemporary architecture and city planning. The life of living structure lies on the smallest scales or fine structure (Alexander 2002–2005) as demonstrated in Figures 1 and 2. It is time to change our mindsets toward fractal geometry, power-law statistics, and Alexandrian living geometry to develop sustainable cities and architecture.

## 5. Concluding remarks

A city is not a simple network, as simple as a regular or random network. Instead, a city is a complex network, or a middle status between the regular and random counterparts. It is highly efficient locally and globally, inherited respectively from regular and random counterparts. Many urban theories, such as the Central Place Theory, Zipf's Law, the Image of the City, and the Theory of Centers can be better understood from the perspective of complex networks. Network cities bear the scaling hierarchy of far more small things than large ones, or living structure in general. This is the source of structural beauty and the image of the city. The scaling hierarchy should be interpreted more broadly, i.e., far more unpopular things than popular ones in terms of topology, or far more meaningless things than meaningful ones in terms of semantics. In this connection, a city is indeed a tree in terms of the scaling hierarchy.

The kind of complex network thinking is manifested in a series of Alexander's works that are highly iterative, such as *Notes on the Synthesis of Form* (Alexander 1964), *A City Is not a Tree* (Alexander 1965), *The Timeless Way of Building* (Alexander 1979), and *The Nature of Order* (Alexander 2002–2005). The complex-network perspective implies that within a city, every element depends on every other element, and changing one element would affect virtually every other in a design context. In this chapter, I have shown the power of complex-network perspective in understanding city complexity, in particular the topological view of city structure. Further work is expected towards the integration of the Theory of Centers and network science, and of living geometry and fractal geometry, for sustainable urban design.

**References:**
Alexander C. (1964), *Notes on the Synthesis of Form*, Harvard University Press: Cambridge, Massachusetts.
Alexander C. (1965), A city is not a tree, *Architectural Forum*, 122(1+2), 58-62.
Alexander C. (1979), *The Timeless Way of Building*, Oxford University Press: Oxford.

Alexander C. (2002-2005), *The Nature of Order: An essay on the art of building and the nature of the universe*, Center for Environmental Structure: Berkeley, CA.

Barabási A. and Albert R. (1999), Emergence of scaling in random networks, *Science*, 286, 509–512.

Christaller W. (1933, 1966), *Central Places in Southern Germany*, Prentice Hall: Englewood Cliffs, N.J.

Köhler W. (1947), *Gestalt Psychology: An introduction to new concepts in modern psychology*, LIVERIGHT: New York.

Jiang B. (2013a), Head/tail breaks: A new classification scheme for data with a heavy-tailed distribution, *The Professional Geographer*, 65 (3), 482 – 494.

Jiang B. (2013b), The image of the city out of the underlying scaling of city artifacts or locations, *Annals of the Association of American Geographers*, 103(6), 1552-1566.

Jiang B. (2015a), Head/tail breaks for visualization of city structure and dynamics, *Cities*, 43, 69-77.

Jiang B. (2015b), Wholeness as a hierarchical graph to capture the nature of space, *International Journal of Geographical Information Science*, 29(9), 1632–1648.

Jiang B., Zhao S., and Yin J. (2008), Self-organized natural roads for predicting traffic flow: a sensitivity study, *Journal of Statistical Mechanics: Theory and Experiment*, July, P07008.

Jiang B. and Claramunt C. (2004), Topological analysis of urban street networks, *Environment and Planning B: Planning and Design*, 31(1), 151-162.

Jacobs J. (1961), *The Death and Life of Great American Cities*, Random House: New York.

Lynch K. (1960), *The Image of the City*, The MIT Press: Cambridge, Massachusetts.

Mandelbrot B. (1982), *The Fractal Geometry of Nature*, W. H. Freeman and Co.: New York.

Mehaffy M. W. and Salingaros N. A. (2006), Geometrical fundamentalism, In: Salingaros N. A. (2006), *A Theory of Architecture*, Umbau-Verlag: Solingen.

Mehaffy M. W. and Salingaros N. A. (2015), *Design for a Living Planet: Settlement, science, and the human future*, Sustasis Press: Portland, Oregon.

Newman M., Barabási A.-L., Watts D. J. (2006, editors), *The Structure and Dynamics of Networks*, Princeton University Press: Princeton, N.J.

Salingaros N. A. (2005), *Principles of Urban Structure*, Techne: Delft.

Watts D. J. and Strogatz S. H. (1998), Collective dynamics of `small-world' networks, *Nature*, 393, 440-442.

Zipf G. K. (1949), *Human Behaviour and the Principles of Least Effort*, Addison Wesley: Cambridge, MA.